\documentclass[pre,superscriptaddress,tightenlines,showpacs,nofootinbib,notitlepage]{revtex4-2}
\usepackage{graphicx,amssymb,amsmath,epsf,bm}
\usepackage[normalem]{ulem}
\usepackage{stackrel}
\usepackage{color}
\usepackage{upgreek}

\DeclareFontFamily{OT1}{pzc}{}
\DeclareFontShape{OT1}{pzc}{m}{it}{<-> s * [1.10] pzcmi7t}{}
\DeclareMathAlphabet{\mathpzc}{OT1}{pzc}{m}{it}

\usepackage{latexsym,amsmath,amsfonts,tabularx,amssymb,bm,empheq}
\usepackage{graphicx,epstopdf}
\usepackage{xspace}

\usepackage{hyperref}
\hypersetup{
  citecolor = {green},
  colorlinks = {true}, 
  urlcolor = {blue} 
}

\newcommand{\bea}{\begin{eqnarray}}
\newcommand{\eea}{\end{eqnarray}}

\def\simge{\mathrel{%
   \rlap{\raise 0.511ex \hbox{$>$}}{\lower 0.511ex \hbox{$\sim$}}}}
\def\simle{\mathrel{
   \rlap{\raise 0.511ex \hbox{$<$}}{\lower 0.511ex \hbox{$\sim$}}}}

\def\simle{\mathrel{
   \rlap{\raise 0.511ex \hbox{$<$}}{\lower 0.511ex \hbox{$\sim$}}}}

\def\simge{\mathrel{%
    \rlap{\raise 0.511ex \hbox{$>$}}{\lower 0.511ex \hbox{$\sim$}}}}

\usepackage{subfigure}
\usepackage{color}
\usepackage{tabularx}

\definecolor{CornFlowerBlue}{rgb}{0.39, 0.58, 0.93}
\definecolor{ForestGreen}{rgb}{0.13, 0.55, 0.13}
\definecolor{WildStrawberry}{rgb}{1.0, 0.26, 0.64}

\usepackage{cleveref}

\begin{document}

\title{Constraint correlation functions of the one-dimensional Ising model in the scaling limit}

\author{Ivan Balog}
 \affiliation{Institute of Physics, Bijeni\v{c}ka cesta 46, HR-10001 Zagreb, Croatia}
\author{Adam Ran\c con}
\affiliation{Univ. Lille, CNRS, UMR 8523 -- PhLAM -- Laboratoire de Physique
des Lasers Atomes et Mol\'ecules, F-59000 Lille, France}
\affiliation{Institut Universitaire de France}


\begin{abstract}
We study the correlation function of the one-dimensional Ising model at fixed magnetization. Focusing on the scaling limit close to the zero-temperature fixed point, we show that this correlation function, in momentum space,  exhibits surprising oscillations as a function of the magnetization. We show that these oscillations have a period inversely proportional to the momentum and give an interpretation in terms of domain walls. This is in sharp contrast with the behavior of the correlation function in constant magnetic fields, and sheds light on recent results obtained by Monte Carlo simulations for the correlation functions of the critical two-dimensional Ising model at fixed magnetization.
\end{abstract}

\maketitle


\section{Introduction}

The Ising model is a cornerstone of statistical physics, thanks to its ability to capture the most stringent aspects of phase transitions despite its apparent simplicity. Of foremost importance,  it can be solved exactly for arbitrary temperature and zero magnetic field in two dimensions (2D), and arbitrary temperature and (constant) field in one dimension (1D) \cite{McCoyWu}. The latter case only displays an ordering transition at zero temperature but is nevertheless of interest, as it is governed at low temperature by localized excitations (domain walls, kinks or instantons). Those excitations, appearing in a variety of classical and quantum problems, are typically non-perturbative and are usually treated using ad hoc methods, e.g. introducing droplets \cite{Bruce1981,Bruce1983} or using instantonic calculations \cite{Rulquin2016}. This calls for developing more generic methods, for instance, building on approximation schemes of the Functional Renormalization Group (FRG) \cite{Dupuis2021}, as was undertaken in \cite{Farkas2023,balog_25_droplet}. Understanding the relation between FRG and exact results for the 1D Ising model close to its zero-temperature transition is thus of interest, to better understand how such excitations can be identified in FRG calculations.

Another interesting aspect of the Ising model is that it allows for studying the emergence of collective phenomena that take place when the correlation length $\xi$ (of a hypothetical infinite system) is much larger than the actual system size $N$. This happens for instance close to a second-order phase transition where universality is expected to appear. One such universal property is given by the probability distribution function (PDF) of the magnetization, i.e. the sum of all the spins, which becomes non-Gaussian due to the strong correlations between the spins, yet universal and depending only on the ratio $N/\xi$ and the boundary conditions. The PDF of the Ising universality class has been studied in 2D and 3D both numerically \cite{Binder1981,Binder1981a,Bruce1992,Nicolaides1988,Tsypin1994,Tsypin2000,Malakis2014,Xu2020} and using perturbative RG approaches \cite{Bruce1979,Eisenriegler1987,Esser1995,Bruce1997,Rudnick1998,sahu_generalization_2025,sahu_probability_2025}, while that of the 1D Ising model has been computed for different boundary conditions using various methods \cite{Antal2004,Garcia-Pelayo2009,Wang2017,Dantchev2022,Dantchev2023,Dantchev2024,stepanyan_thermal_2024}. Recently, a FRG implementation for the PDF's calculation has been devised \cite{Balog2022,balog_universal_2025,rancon_probability_2025}. In this framework, a functional equation is solved to obtain the log of the PDF, called the rate function in probability theory, and the constraint effective potential in field theory.  This equation is naturally expressed in terms of the ``constraint correlation functions'', i.e. the correlation functions between spins at fixed magnetization. We have studied these correlation functions at the critical point of the 2D and 3D Ising model both with FRG and Monte Carlo simulations in \cite{Rose2025}. There, we observed in our 2D Monte Carlo data surprising oscillations in the magnetization dependence of the constraint correlation functions (in momentum space and with periodic boundary conditions). 

To better understand the origin of this non-trivial behavior, we study here the constraint correlation function of the 1D Ising model in the scaling regime close to its zero-temperature critical point. In the limit $N/\xi\to 0$, we do find the same kind of oscillations. We show that they are induced by the imposed magnetization, which creates a new length scale associated with the typical size of the spin domains. This is in sharp contrast with the correlation function in constant field.

The manuscript is organized as follows. In Sec.~\ref{sec:const} we define and compute the PDF and the constraint correlation function for the finite-size 1D Ising model.  In Sec.~\ref{sec:scal} we compute the scaling limit of the constraint correlation function, and analyze its non-trivial behavior as a function of the magnetization. Finally, in Sec.~\ref{sec:disc}, we discuss our results and in particular their implications for functional methods such as FRG. Some technical details about the calculations are given in the appendices.

\section{Finite-size constraint correlation function \label{sec:const}}

We are interested in the correlation function of the 1D Ising model in the ``canonical'' ensemble with fixed magnetization $\sum_i \sigma_i = S$. Here and below, the sum is over the $N$ sites of the lattice, and we assume periodic boundary conditions. It is useful to first define the constraint partition function 
\begin{equation}
    Z_N(S) = \langle \delta_{\sum_i\sigma_i,S} \rangle,
\end{equation}
where the thermal average is over the usual Boltzmann weight
\begin{equation}
   \langle \ldots \rangle= \sum_{\{\sigma_i\}} \ldots e^{-H[\sigma,0]}/\mathcal Z_N(0).
\end{equation}
The Hamiltonian is that of the ferromagnetic nearest-neighbor 1D Ising model in a magnetic field
\begin{equation}
   H[\sigma,h] = -K\sum_i \sigma_i\sigma_{i+1}-h\sum_i \sigma_i,
\end{equation}
and $\mathcal Z_N(h)=\sum_{\{\sigma_i\}} e^{-H[\sigma,h]}$ is the (unconstrained) partition function.
Of course, $Z_N(S)=P_N(S)$, the PDF of the magnetization that has been studied extensively in  \cite{Antal2004,Garcia-Pelayo2009,Wang2017,Dantchev2022,Dantchev2023,Dantchev2024,stepanyan_thermal_2024}. The magnetization $S$ can only take $N$ possible values, from $-N$ to $N$ with increments of $2$ (at fixed $N$, the PDF, seen as a function of the real variable $S$, vanishes for all other values of $S$). Furthermore, the PDF is normalized, i.e. $\sum_S Z_N(S)=1$.

The constraint correlation function between two spins at position $i$ and $j$, the main object of our study, is defined as 
\begin{equation}
    G_N(r;S) = \langle \sigma_i\sigma_j\delta_{\sum_i\sigma_i,S} \rangle/Z_N(S),
    \label{eq:Gconstr}
\end{equation}
which is a function of $r=|i-j|$ by invariance by translation induced by the periodic boundary conditions.
Note that it obeys the sum rule $\sum_{r=1}^N G_N(r;S) = S^2/N$, since $\langle \sigma_j\delta_{\sum_i\sigma_i,S} \rangle/Z_N(S)=S/N$ by invariance by translation.

The constraint partition function as well as $\mathbf{G}_N(r;S)=\langle \sigma_i\sigma_j\delta_{\sum_i\sigma_i,S} \rangle$ can be computed using a transfer matrix approach as follows, see \cite{Wang2017,Dantchev2022,Dantchev2023,stepanyan_thermal_2024} for similar calculations of the former. First, we exponentiate the constraint using $\delta_{n,m}=\int_{-\pi}^\pi \frac{d\theta}{2\pi}e^{i(n-m)}$, and the constraint partition function can be rewritten as
\begin{equation}
    Z_N(S) = \int_{-\pi}^\pi \frac{d\theta}{2\pi}e^{-i S\theta}\langle e^{i\theta \sum_i \sigma_i} \rangle=\int_{-\pi}^\pi \frac{d\theta}{2\pi} e^{-i S\theta}\frac{\mathcal Z_N(h=i\theta)}{\mathcal Z_N(0)}.
    \label{eq:constraintZ}
\end{equation}

Both $Z_N(S)$ and $G_N(r;S)$ can be computed from the unconstrained partition function and correlation function, and we recall now these standard results, see \cite[Chap. III]{McCoyWu} for an introduction to the method.
The transfer matrix is 
\begin{equation}
    T(h)=\begin{pmatrix}
        e^{K+h} && e^{-K}\\
        e^{-K} && e^{K-h}
    \end{pmatrix},
\end{equation}
with eigenvalues 
\begin{equation}
        \lambda_\pm(h) = e^{K}\left(\cosh(h) \pm \sqrt{e^{-4K}+\sinh^2(h)}\right),
\label{eq:lambdapm}
\end{equation}
and eigenvectors 
\begin{equation}
\begin{split}
            u_+&=\begin{pmatrix}
            \cos(\phi/2)\\
            \sin(\phi/2)
        \end{pmatrix},\\
        u_-&=\begin{pmatrix}
           - \sin(\phi/2)\\
            \cos(\phi/2)
        \end{pmatrix},      
\end{split}
\end{equation}
with
\begin{equation}
\begin{split}
            \sin(\phi(h))&=\frac{1}{\sqrt{e^{-4K}+\sinh^2(h)}},\\
        \cos(\phi(h))&=\frac{\sinh(h)}{\sqrt{e^{-4K}+\sinh^2(h)}}.
\end{split}
\label{eq:phi}
\end{equation}
Using the periodic boundary conditions, one finds that
\begin{equation}
\begin{split}
    \mathcal Z_N(h) &= {\rm Tr} \left(T(h)^N\right),\\
                    &= \lambda_+(h)^N + \lambda_-(h)^N.
\end{split}
\end{equation}

Applying the same method for $\mathbf{G}_N(r;S)$, one gets
\begin{equation}
\begin{split}
    \mathbf{G}_N(r;S) &= \int_{-\pi}^\pi \frac{d\theta}{2\pi}e^{-i S\theta}\langle \sigma_i\sigma_j e^{i\theta \sum_i \sigma_i} \rangle,\\
                &=\int_{-\pi}^\pi \frac{d\theta}{2\pi} \mathcal G_N(i,j;i\theta),
\end{split}
\label{eq:GG}
\end{equation}
with $\mathcal G_N(r;h)$ the correlation function in a field $h$. The transfer matrix method gives ($\hat\sigma_z$ the third Pauli matrix),
\begin{equation}
\begin{split}
   \mathcal G_N(r;h) &= {\rm Tr} \left(\hat\sigma_z T(h)^{|i-j|} \hat\sigma_z T(h)^{N-|i-j|}\right)/\mathcal Z_N(0),\\
                &= \frac{\lambda_+(h)^{N-r}\lambda_-(h)^{r}+\lambda_-(h)^{N-r}\lambda_+(h)^{r}}{\lambda_+(0)^N + \lambda_-(0)^N}\sin^2(\phi(h))
                +\frac{\lambda_+(h)^{N}+\lambda_-(h)^{N}}{\lambda_+(0)^N + \lambda_-(0)^N}\cos^2(\phi(h)).
\end{split}
\end{equation}

In zero field and in the thermodynamic limit $N\to\infty$, using $\lambda_+>\lambda_-$, we recover the standard result for the partition function $\mathcal Z_N(0)\to 2^N \cosh^N(K)$ and the correlation function
\begin{equation}
\begin{split}
    \mathcal G(r;0)&\to \left(\tanh(K)\right)^{r}= e^{-r/\xi},
\end{split}
\end{equation}
with $\xi=-1/\log(\tanh(K))$ the correlation length of the infinite system.

\section{Scaling limit \label{sec:scal}}

We are interested in the scaling limit where the system size $N$ becomes infinite while keeping the ratio $\zeta=N/2\xi$ constant. This amounts to properly take the limit $K\to\infty$, with $\xi\simeq e^{2K}/2$ in this limit, i.e. we take the limit $N\to\infty$, $K\to\infty$, such that $Ne^{-2K}=\zeta$ \cite{Antal2004}.

In this limit, the zero-field partition function takes the form $\mathcal{Z}_N(0)\to e^{NK}2\cosh(\zeta)$, while the zero-field correlation function becomes ($r/N=x$ fixed)
\begin{equation}
    \mathcal G_N(r;0)\to \mathpzc g_\zeta(x;0)= \frac{\cosh((1-2x)\zeta)}{\cosh(\zeta)}.
    \label{eq:standardGscal}
\end{equation}
The magnetic susceptibility $\chi=\frac{1}{N}\langle(\sum_i\sigma_i)^2\rangle$ reads ($v=\tanh (K)$)
\begin{equation}
\begin{split}
    \chi = \frac{1+v}{1-v}\frac{1-v^N}{1+v^N}\to N\frac{\tanh( \zeta)}{\zeta}.
\end{split}
\end{equation}
Recalling that for critical systems, in the scaling limit, the finite-size scaling form of the susceptibility is
\begin{equation}
    \chi=N^{2-\eta}f_\chi(N/\xi),
\end{equation}
we infer that $\eta=1$, as can be deduced from droplets analyses \cite{Bruce1981,Bruce1983}.

This implies that the typical fluctuations of the magnetization are of order one, as follows from its scaling $N^{\frac{d-2+\eta}{2}}=1$, while the corresponding typical magnetic field is of order $N^{-\frac{d+2-\eta}{2}}=N^{-1}$.
The correct scaling limit of the field-dependent partition function is therefore
\begin{equation}
    \begin{split}
        \mathcal{Z}_N(h)\to \mathpzc z_\zeta(x;\tilde h)=e^{NK}2 \cosh\left(\sqrt{\zeta^2+\tilde h^2}\right),
    \end{split}
    \label{eq:scalZ}
\end{equation}
with $\tilde h=hN$ of order 1, and where we have used that in the scaling limit,
\begin{equation}
    \begin{split}
        \lambda_\pm(h)^N&=e^{NK}\left(\cosh(h) \pm \sqrt{e^{-4K}+\sinh^2(h)}\right)^N,\\
        &\simeq e^{NK}\left(1 \pm \sqrt{\zeta^2+(hN)^2}/N\right)^N,\\
        &\to e^{NK\pm \sqrt{\zeta^2+(hN)^2}}.
    \end{split}
    \label{eq:scal_lambda}
\end{equation}

We are now in a position to compute the constraint partition function and correlation function in the scaling limit (see App.~\ref{app:calc} for more details about the following calculations). 

\subsection{Constraint partition function}

The constraint partition function Eq.~\eqref{eq:constraintZ} can be rewritten as
\begin{equation}
    \begin{split}
    Z_N(S) & = 2\int_{-\frac\pi2}^{\frac\pi2}\frac{d\theta}{2\pi} \cos(S \theta) \frac{\mathcal Z_N(i\theta)}{Z_N(0)},
    \end{split}
\end{equation}
using the symmetry of the integrand.

In the scaling limit, $N\to\infty$, $Ne^{-2K}=\zeta$, the magnetization $s=S/N$ becomes a continuous variable in $[-1,1]$, and the magnetization spacing $2/N\to ds$, such that $Z_N(S)\to z_\zeta(s)ds$ with
\begin{equation}
    \begin{split}
    z_\zeta(s)&=\lim_{N\to \infty} \frac{N}{2}Z_N(S),\\
        &= \lim_{N\to \infty} N\int_{-\frac\pi2}^{\frac\pi2}\frac{d\theta}{2\pi} \cos(S \theta) \frac{\mathcal Z_N(i\theta)}{Z_N(0)},\\
        &\to\int_{-\infty}^{\infty} \frac{dy}{2\pi} \frac{\cosh(\sqrt{\zeta^2-y^2})}{\cosh(\zeta)}\cos(sy).
    \end{split}
\end{equation}
Here we have performed the change of variable $y=N\theta$ and used the scaling limit, Eq.~\eqref{eq:scalZ}.
Performing the integral over $y$ gives
\begin{equation}
    \begin{split}
    z_\zeta(s)&=\frac{\zeta}{2\sqrt{1-s^2}}\frac{I_1(\zeta\sqrt{1-s^2})}{\cosh(\zeta)}+\frac{\delta(s-1)+\delta(s+1)}{2\cosh(\zeta)},
    \end{split}
\label{eq:Zconstscal}
\end{equation}
with $I_\nu(z)$ the modified Bessel of the first kind, which is exactly the result obtained in \cite{Antal2004} by a combinatorial method.

\subsection{Constraint correlation function}

Using the symmetries of the integral, we can rewrite Eq.~\eqref{eq:GG} as
\begin{equation}
\begin{split}
    \mathbf{G}_N(r;S) =2\int_{-\pi/2}^{\pi/2} \frac{d\theta}{2\pi} \bigg(&\frac{\lambda_+(i\theta)^{N-r}\lambda_-(i\theta)^{r}+\lambda_-(i\theta)^{N-r}\lambda_+(i\theta)^{r}}{\mathcal Z_N(0)}\sin^2(\phi(i\theta)) \\
         &+\frac{\lambda_+(i\theta)^{N}+\lambda_-(i\theta)^{N}}{\mathcal Z_N(0)}\cos^2(\phi(i\theta))\bigg)\cos(S\theta),
\end{split}
\label{eq:GGtheta}
\end{equation}
The scaling limit is taken as $N\to\infty$, at fixed $Ne^{-2K}=\zeta$, $s=S/N$ and $x=r/N\in[0,1]$, with $ \mathbf{G}_N(r;S)\to \mathbf{g}_\zeta(x;s)ds$.
Using the change of variable $y=N\theta$, Eq.~\eqref{eq:scal_lambda} as well as $\sin^2(\phi(i\theta))\to\zeta^2/(\zeta^2-y^2)$ and $\cos^2(\phi(i\theta))\to-y^2/(\zeta^2-y^2)$, we obtain
\begin{equation}
\begin{split}
    \mathbf{g}_\zeta(x;s) = \int_{-\infty}^{\infty}\frac{dy}{2\pi}\cos(sy) \left(\frac{\zeta^2}{\zeta^2-y^2}\frac{\cosh\left((1-2x)\sqrt{\zeta^2-y^2}\right)}{\cosh(\zeta)}-\frac{y^2}{\zeta^2-y^2}\frac{\cosh(\sqrt{\zeta^2-y^2})}{\cosh(\zeta)}\right).
\end{split}
\end{equation}
One can show that $\int_0^1dx \,\mathbf{g}_\zeta(x;s)=s^2 z_\zeta(s)$ as expected from the sum rule.

We are now in a position to compute the constraint correlation function in momentum space. Due to the periodic boundary conditions, momenta at finite size are of the form $2\pi n/N$, $n\in\{1,\ldots,N\}$, and in the scaling limit $ 2\pi n r/N\to 2\pi n x$. Thus, we need to compute $\mathbf{g}_\zeta(n;s)=\int_0^1 dx\, 2\cos(2\pi n x) \mathbf{g}_\zeta(x;s)$, i.e. (here and below, $n\neq 0$)
\begin{equation}
\begin{split}
    \mathbf{g}_\zeta(n;s) &= \int_0^1 dx \int_{-\infty}^{\infty}\frac{dy}{2\pi} \frac{\zeta^2}{\zeta^2-y^2}\frac{\cosh((1-2x)\sqrt{\zeta^2-y^2})}{\cosh(\zeta)}2\cos(2\pi n x)\cos(sy),\\
                    &= \frac{\zeta^2}{\cosh(\zeta)}\int_{|s|}^1 dt\frac{\sin((1-t)n\pi)}{n\pi} I_0(\zeta\sqrt{t^2-s^2}).
\end{split}
\end{equation}

Finally, we obtain our main result, the scaling limit of the constraint correlation function Eq.~\eqref{eq:Gconstr}, in momentum space, $g_\zeta(n;s)=\mathbf{g}_\zeta(n;s)/z(s)$,
\begin{equation}
\begin{split}
    g_\zeta(n;s) = 2\zeta\sqrt{1-s^2}\int_{|s|}^1 dt\frac{\sin((1-t)n\pi)}{n\pi} \frac{I_0(\zeta\sqrt{t^2-s^2})}{I_1(\zeta\sqrt{1-s^2})}.
\end{split}
\end{equation}

\begin{figure}
    \centering
    \includegraphics[width=0.5\linewidth]{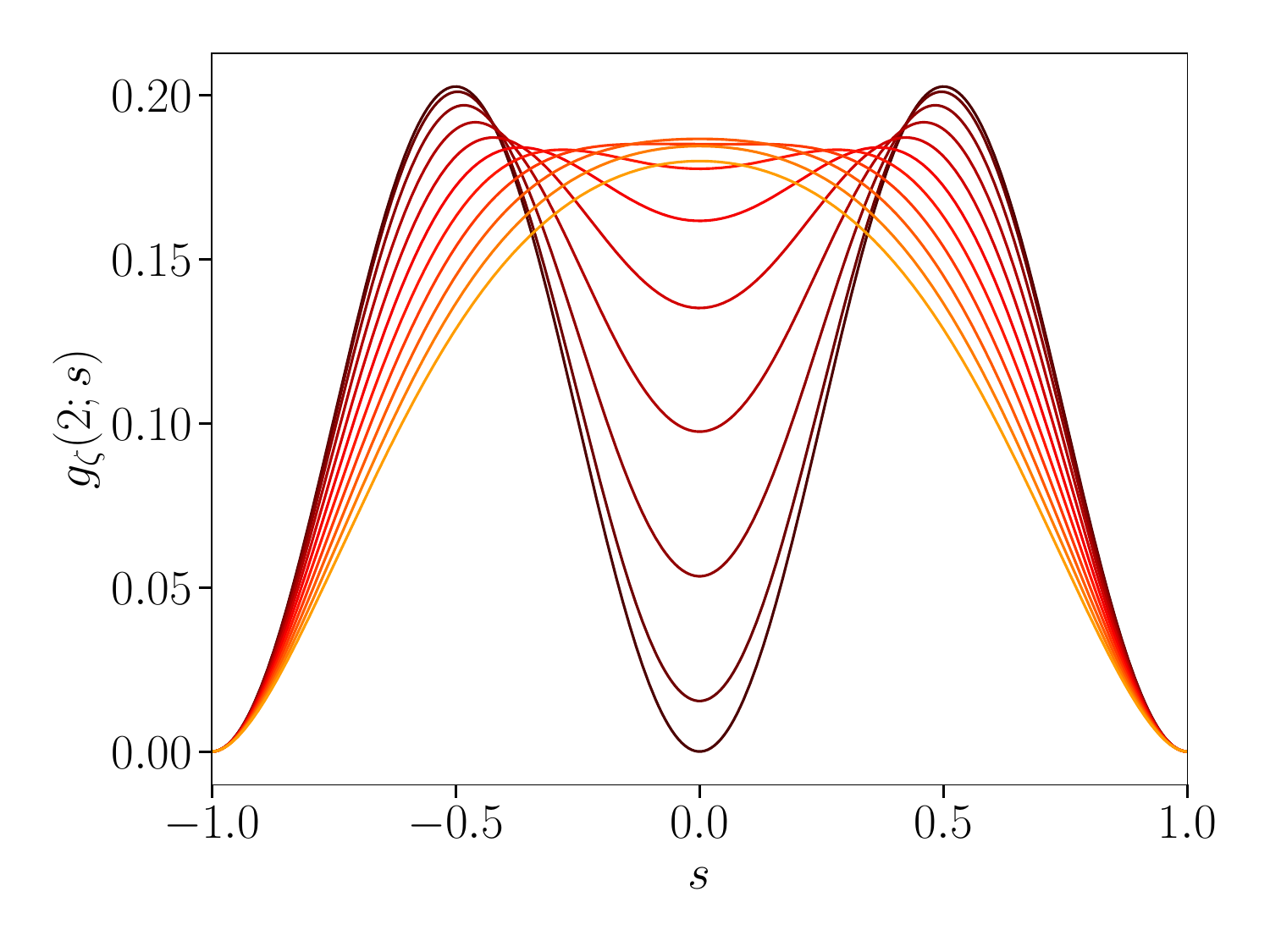}
    \caption{Constraint correlation function in the scaling limit as a function of $s$ at fixed $n=2$  for increasing values of $\zeta\in[0,8]$ (from dark to light colors, same legend as Fig.~\ref{fig:g_x}). }
    \label{fig:g_zeta}
\end{figure}

\begin{figure}
    \centering
    \includegraphics[width=0.5\linewidth]{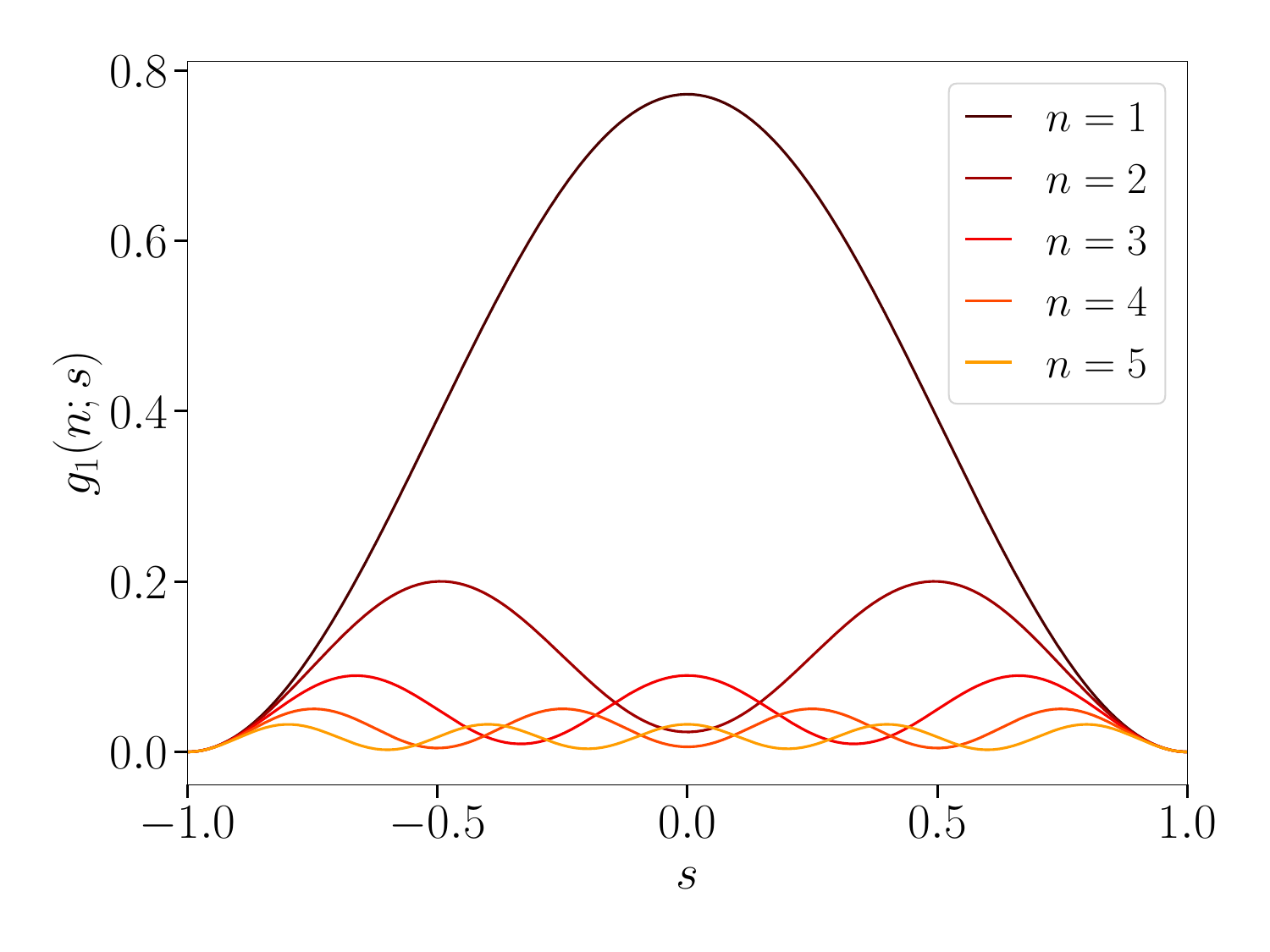}
    \caption{Constraint correlation function in the scaling limit as a function of $s$ at fixed $\zeta=1$  for increasing values of $n\in\{1,2,3,4,5\}$ (from dark to light colors). }
    \label{fig:g_n}
\end{figure}

The behavior of $g_\zeta(n;s)$ as a function of  $s$ for various $\zeta$ at fixed $n=2$ is shown in Fig.~\ref{fig:g_zeta}. We observe that for $\zeta$ small enough ($\zeta\lesssim 6$ for $n=2$), i.e. when the (infinite system) correlation length becomes of the order of or much larger than the system size, the constraint correlation function shows oscillations in $s$ for a fixed $n$. The period of those oscillations depends on $n$ as can be seen in Fig.~\ref{fig:g_n} where we fix $\zeta=1$. This behavior can be captured analytically in the limit $\zeta\to0$, where we find that
\begin{equation}
\begin{split}
    g_0(n;s)=\frac{4 (1-\cos (\pi  n (1-|s|)))}{\pi ^2 n^2},
\end{split}
\label{eq:gscal}
\end{equation}
has a period of $2/n$ in $s$.

\begin{figure}
    \centering
    \includegraphics[width=0.5\linewidth]{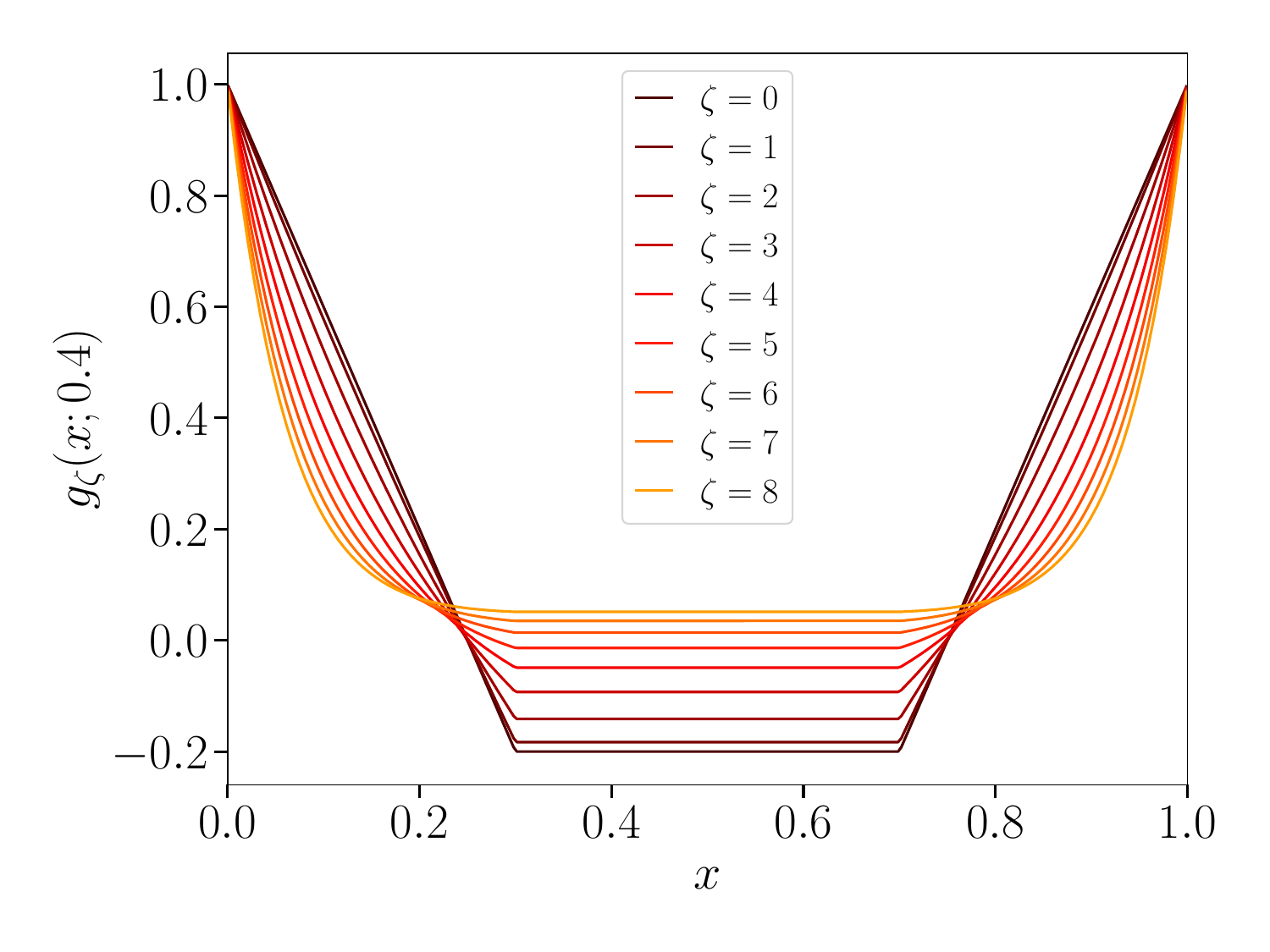}
    \caption{Constraint correlation function in real space in the scaling limit as a function of $x$ at fixed magnetization $s=0.4$  for increasing values of $\zeta\in[0,8]$ (from dark to light colors). }
    \label{fig:g_x}
\end{figure}

At first sight, these oscillations might appear surprising, since the only macroscopic scale, the correlation length $\xi$ is much larger than the system size. This, however, overlooks a new scale introduced by the constraint, which is given by $s$. 
Indeed, in the limit $\zeta\to0$, the temperature is so small that the only thermally allowed states, if there is no constraint on $s$, are the two completely polarized states. These states have all spins either up or down, corresponding to $|s|=1$. Now, the constraint $|s|<1$ implies the existence of two distinct domains. One domain has a density of $(1+s)/2$ up spins, while the other has a density of $(1-s)/2$ down spins. This condition creates a new scale related to the size of the domains, which is of the order $1-|s|$ (in units of the system size). This new scale allows for the non-trivial structure of the constraint correlation function.
This can be checked by an explicit calculation the correlation function in the presence of only two domains, see App.~\ref{app:domain}, which reads in the scaling limit
\begin{equation}
\begin{split}
    g_0(x;s)=1-2\,{\rm max}(|s|,|1-2x|)
\end{split}
\label{eq:g2domain}
\end{equation}
the Fourier transform of which is Eq.~\eqref{eq:gscal}. The oscillating term $\cos (\pi  n (1-|s|))$ is the signature in Fourier space of the discontinuous derivatives at $x=(1\pm s)/2$.

The real-space correlation function can be expressed as~\footnote{The integral can be performed explicitly at least in the case $s=0$, and is given by a cumbersome combination of special functions, which we do not give.}
\begin{equation}
    g_\zeta(x;s)=1-\zeta\sqrt{1-s^2}\int_{{\rm max}(|s|,|1-2x|)}^1 \frac{I_0(\zeta\sqrt{t^2-s^2})}{I_1(\zeta\sqrt{1-s^2})} dt .
\end{equation}
This expression gives back Eq.~\eqref{eq:g2domain} in the limit $\zeta\to 0$, and satisfies explicitly $g_\zeta(0;s)=1=g_\zeta(1;s)$, imposed by $\hat\sigma_i^2=1$. It is shown for a representative value of the magnetization $s=0.4$ for various values of $\zeta$ in Fig.~\ref{fig:g_x}. It decreases for increasing $x\in [0,(1-|s|)/2]$, the remains constant up to $(1+|s|)/2$, before increasing again due to the periodic boundary condition and the parity symmetry. At small $\zeta$, the discontinuous derivatives at $x=\dfrac{1\pm s}{2}$ are clearly visible, but become smaller at higher temperature (the derivatives jump scales as $\exp(-\sqrt{1-s^2}\zeta)$ at large $\zeta$).

Therefore, the reason why the oscillations of the correlation function in momentum space disappear when $\zeta$ (i.e. the temperature) is increased is clear: As one reduces the (infinite system) correlation length, the spin-up and spin-down domains can be split into many smaller domains which can be placed in many more places, thus blurring the sharp behavior of $g_\zeta(x;s)$.

\begin{figure}
    \centering
    \includegraphics[width=0.5\linewidth]{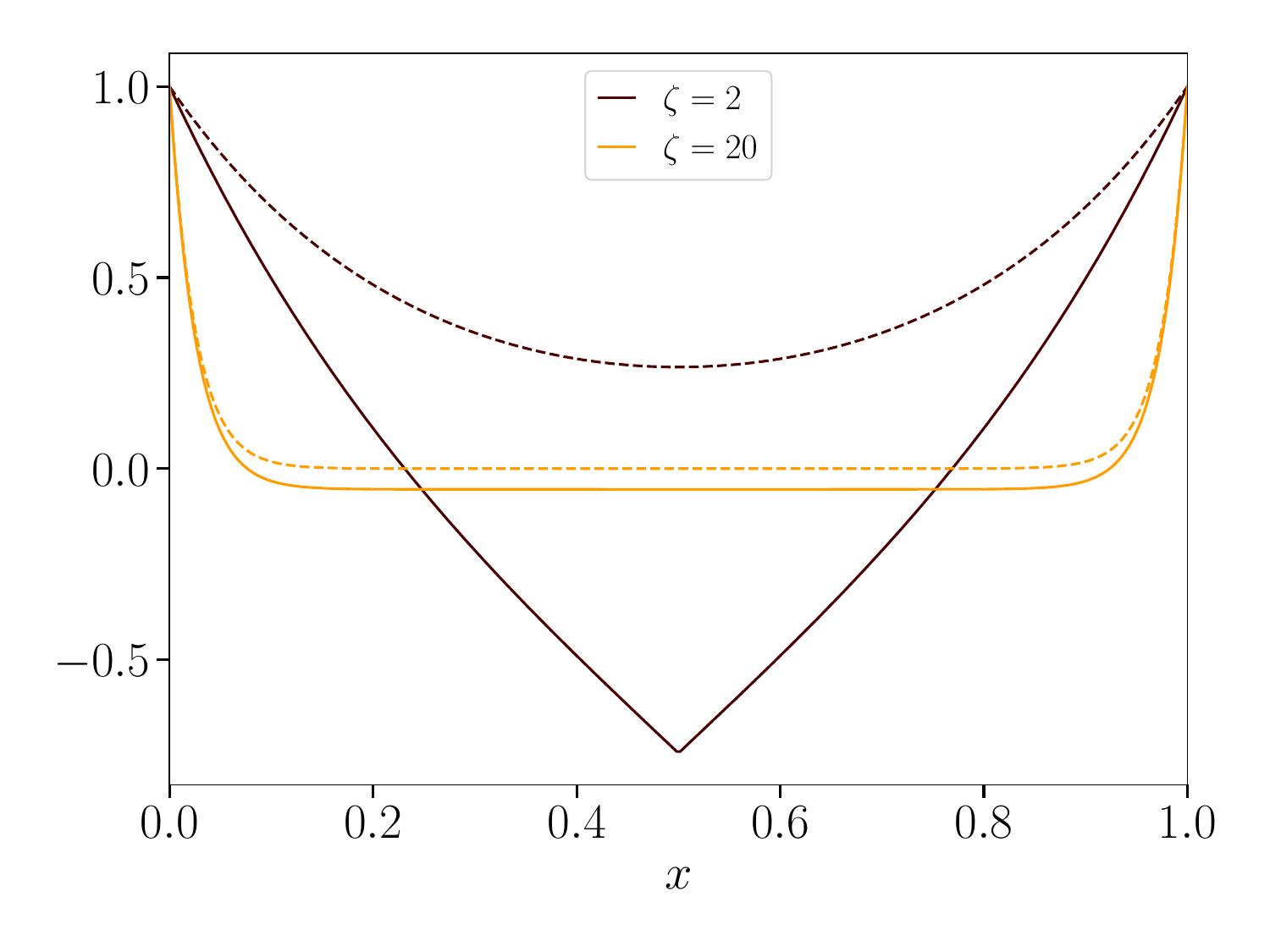}
    \caption{Comparison between the constraint (at $s=0$, full lines) and unconstrained (at $m=0$, dashed lines) correlation functions in real space in the scaling limit as a function of $x$ at small and large $\zeta$. }
    \label{fig:comp_g_x}
\end{figure}

\section{Discussion \label{sec:disc}}

We have computed the constraint correlation function of the 1D Ising model in the scaling limit. We have shown that when the correlation length is of the order or much larger than the size of the system, marked oscillations in the magnetization appear. This is to be contrasted with the field dependence of the correlation function at fixed magnetic field $h$ or fixed \emph{average} magnetization $m=\langle s\rangle$.
Indeed, the former reads in real space (with $\tilde h=hN$ held constant)
\begin{equation}
\mathpzc g_\zeta(x;\tilde h)=\frac{\zeta ^2} {\zeta ^2+\tilde h^2}\frac{\cosh \left((1-2 x) \sqrt{\zeta
   ^2+\tilde h^2}\right)}{\cosh \left(\sqrt{\zeta
   ^2+\tilde h^2}\right)}+\frac{\tilde h^2}{\zeta ^2+\tilde h^2},
\end{equation}
and in momentum space
\begin{equation}
\mathpzc g_\zeta(n;\tilde h)= (2-\delta_{n,0})\frac{\zeta^2}{\sqrt{\zeta^2+\tilde h^2}}\frac{\tanh\left(\sqrt{\zeta^2+\tilde h^2}\right)}{(n\pi)^2+\zeta^2+\tilde h^2}+\delta_{n,0}\frac{\tilde h^2}{\zeta ^2+\tilde h^2},
\end{equation}
which can be expressed in terms of $m=\frac{\tilde h}{ \sqrt{\zeta^2+\tilde h^2}}\tanh\left(\sqrt{\zeta^2+\tilde h^2}\right)$. Its momentum and field dependence are rather standard and do not show oscillations, since a fixed magnetic field, and thus a fixed average magnetization, only reduces the correlation length. It has been remarked in \cite{gross_statistical_2017} that the constraint implies anti-correlation between spins at large distances to satisfy the sum rule, and that at the mean-field level, both the constrained and unconstrained correlation functions have a similar decay as a function of the distance. Figure \ref{fig:comp_g_x} compares both correlation functions for $s=0=m$, for a small and a large value of $\zeta$. In the latter case, the exponential decay is the same, with the constraint correlation function becoming negative to satisfy the sum rule. For smaller $\zeta$, the two functions are markedly different, with a different decay rate, and a cusp at $x=(1+|s|)/2=1/2$ for the constraint correlation function.

\begin{figure}
    \centering
    \includegraphics[width=0.5\linewidth]{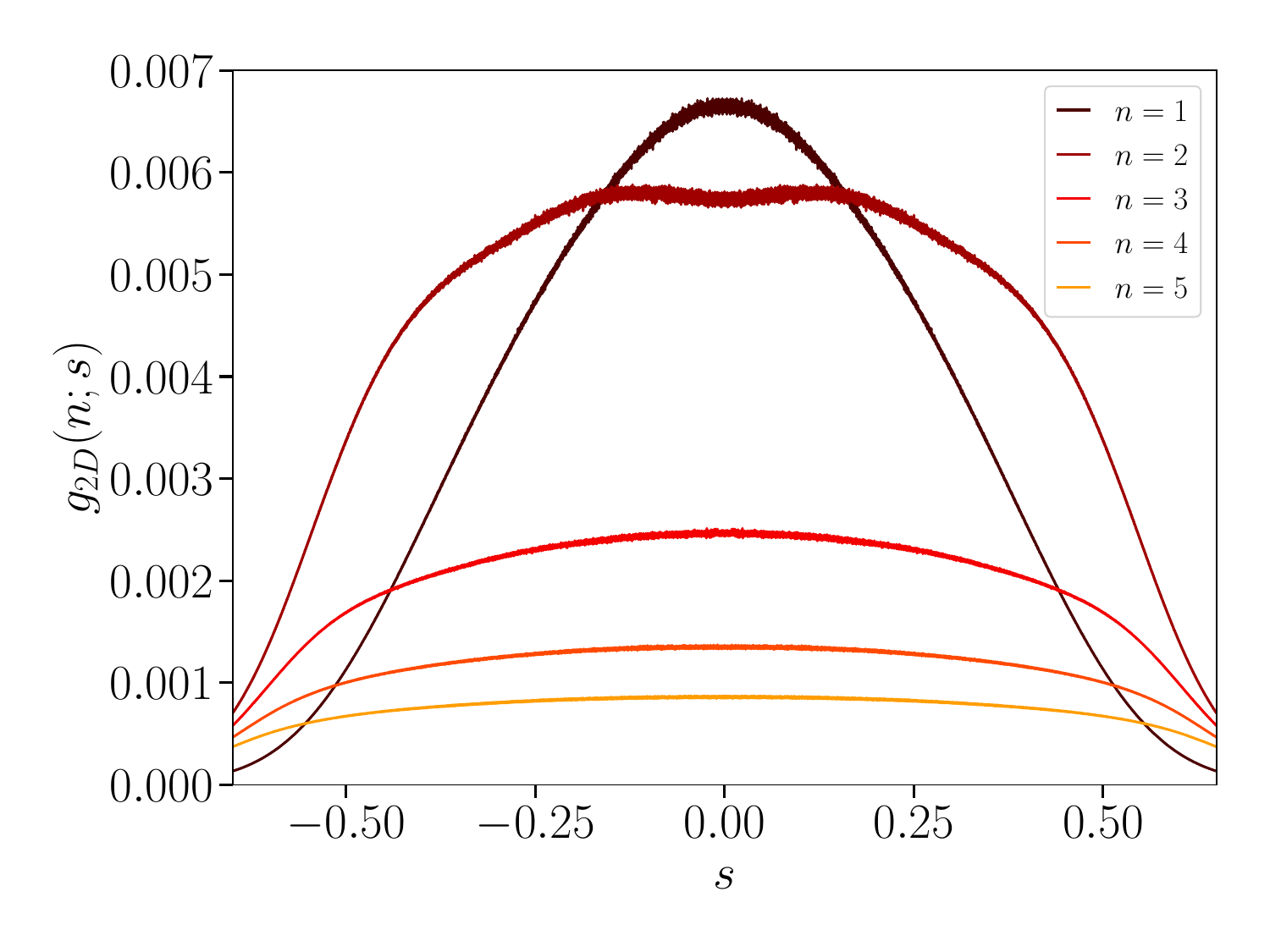}
    \caption{Constraint correlation function of the critical 2D Ising model as a function of $s$ for increasing momenta $(p_x,p_y)=(\frac{2\pi n}{L},0)$ with $n\in \{1,2,3,4,5\}$ (from dark to light colors). The simulations were performed on a square lattice of linear size $L=256$ with nearest-neighbor interactions at $T=T_c$ \cite{Rose2025}.   The correlation function at $n=1$ has been divided by a factor of $6$ for better visibility.}
    \label{fig:2D}
\end{figure}

Let us now come back to the results we have obtained previously in \cite{Rose2025} for the constraint correlation function for the critical 2D Ising model $g_{2D}(n;s)$. The critical constraint correlation function is shown in Fig.~\ref{fig:2D}, obtained at $T=T_c$ (corresponding to $\zeta=0$) for momenta aligned along the $x$-axis. We see that the $n=2$ mode displays an oscillatory behavior reminiscent of that is found in 1D. Note that the typical magnetization over which this behavior happens scales as $L^{-\frac{d-2+\eta}{2}}=L^{-1/8}$ with $L$ the linear size of the system (here $L=256$ for which finite size corrections are negligible), and that the maxima of $g_{2D}(2;s)$ are here at $|s|\simeq 0.1$, very far from the typical magnetization (corresponding to the location of the maximum of the PDF) which is $|s|\simeq 0.6$.

Based on the calculations above, we conclude that this behavior arises from the fixed magnetization, which may introduce a characteristic scale in the critical spin configurations. Unlike the 1D case, this behavior is observed only for $n=2$ and is significantly weaker. It is entirely absent in 3D (not shown). In 2D and 3D, the transition can be described in terms of clusters \cite{FK_72,binder_clusters_1976}, that generalize the notion of correlated domains in higher dimensions. While the fixed magnetization probably also introduces a new scale in the problem, the oscillations might be washed out by the numerous ways the relevant configurations can be placed.

Finally, we would like to comment on the implications of our results on the functional approaches to field theory, such as the FRG. First, standard approximation schemes like the derivative expansion (in which on expands the inverse of the constraint correlation function in momentum) cannot capture the non-trivial dependence in both momentum and magnetization that we have found here. Nevertheless, the derivative expansion near the lower critical dimension succeeds in reproducing several droplet-related features \cite{Farkas2023,balog_25_droplet}. Yet, the relevance of the oscillations for a complete description of droplet physics remains unclear. Identifying FRG schemes capable of incorporating these oscillations may provide a more natural and comprehensive framework for droplet excitations than the derivative expansion.

\acknowledgments 
We wish to acknowledge F\'elix Rose and Bertrand Delamotte for their collaboration on prior projects that led to this work. AR thanks the Institute of Physics of Zagreb, where this work was done, for its hospitality.
This work was supported by the Croatian Science fund project HRZZ-IP-10-2022-9423, an IEA CNRS project, and by the “PHC COGITO” program (project number: 49149VE) funded by the French Ministry for Europe and Foreign Affairs, the French Ministry for Higher Education and Research, and The Croatian Ministry of Science and Education.  IB wishes to acknowledge the support of the INFaR and FrustKor projects financed by the EU through the National Recovery and Resilience Plan (NRRP) 2021-2026. AR has benefited from the financial support of the Grant No. ANR-24-CE30-6695 FUSIoN.

\appendix
\section{Computation of the constraint partition function and correlation function \label{app:calc}}

In this appendix, we give additional details on the calculation of the constraint partition function and correlation function.

\subsection{Constraint partition function}
From Eq.~\eqref{eq:constraintZ}, the constraint partition function is given by 
\begin{equation}
    Z_N(S) = \int_{-\pi}^\pi \frac{d\theta}{2\pi}e^{-i S\theta} \frac{\lambda_+(i\theta)^N+\lambda_-(i\theta)^N}{\lambda_+(0)^N+\lambda_-(0)^N},
    \label{eq:constraintZapp}
\end{equation}
with 
\begin{equation}
        \lambda_\pm(i\theta) = e^{K}\left(\cos(\theta) \pm \sqrt{e^{-4K}-\sin^2(\theta)}\right).
\end{equation}
The integral in Eq.~\eqref{eq:constraintZapp} is invariant under the combined change $\theta\to-\theta$ and $S\to-S$, i.e. $Z_N(S)=Z_N(-S)$, so we can replace $e^{-i S\theta}$ by $\cos(S\theta)$. Furthermore, the integrand is periodic of period $\pi$, as under the change $\theta\to\theta+\pi$, we have $\lambda_\pm(i\theta)\to-\lambda_\mp(i\theta)$, and using $e^{-i S\theta}\to (-1)^Ne^{-i S\theta}$ (since $S=\{-N,-N+2,\ldots,N-2,N\}$).
We can thus rewrite Eq.~\eqref{eq:constraintZapp} as
\begin{equation}
    Z_N(S) = 2\int_{-\pi/2}^{\pi/2} \frac{d\theta}{2\pi}\cos( S\theta) \frac{\lambda_+(i\theta)^N+\lambda_-(i\theta)^N}{\lambda_+(0)^N+\lambda_-(0)^N}.
\end{equation}

In the scaling limit, we must compute $z_\zeta(s)=\frac{N}{2}Z_N(S)$ with $N\to\infty$ keeping $Ne^{-2K}=\zeta$ and $S/N=s$ constant, i.e.
\begin{equation}
    \begin{split}
    z_\zeta(s)     &= \lim_{N\to \infty} N\int_{-\frac\pi2}^{\frac\pi2}\frac{d\theta}{2\pi} \cos(s N \theta) \frac{\left(\cos(\theta) +\sqrt{\zeta^2/N^2-\sin^2(\theta)}\right)^N+\left(\cos(\theta) -\sqrt{\zeta^2/N^2-\sin^2(\theta)}\right)^N}{2\cosh(\zeta)}\cos(s N \theta),\\
        &=\lim_{N\to \infty}\int_{-N\frac\pi2}^{N\frac\pi2}\frac{dy}{2\pi}\frac{\left(\cos(y/N) +\sqrt{\zeta^2/N^2-\sin^2(y/N)}\right)^N+\left(\cos(y/N) -\sqrt{\zeta^2/N^2-\sin^2(y/N)}\right)^N}{2\cosh(\zeta)}\cos(sy),\\
        &=\int_{-\infty}^{\infty} \frac{dy}{2\pi} \frac{\cosh(\sqrt{\zeta^2-y^2})}{\cosh(\zeta)}\cos(sy).
    \end{split}
\end{equation}
The integral over $y$ can be performed as follows. First, we note that $\cosh(\sqrt{\zeta^2-y^2})\simeq \cos(y)$ for $y\gg\zeta$, such that it is convenient to add and subtract $\cos(y)$ in the numerator. Calling 
\begin{equation}
    F(\zeta) = \int_{-\infty}^{\infty} \frac{d\tilde y}{2\pi}\left(\cosh(\zeta\sqrt{1-\tilde y^2})-\cos(\tilde y)\right)\cos(k\tilde y),
\end{equation}
such that  $z_\zeta(s)=\frac{\zeta}{\cosh(\zeta)}F(\zeta)+\int_{-\infty}^{\infty} \frac{dy}{2\pi} \frac{\cos(y)}{\cosh(\zeta)}\cos(sy)$ with $k=s\zeta$ after the change of variable $y=\zeta \tilde y$. Performing the Laplace transform from $F(\zeta)$ to $\hat F(p)$ and computing the integral over $\tilde y$, we find
\begin{equation}
    \hat F(p) = \frac{1}{2}\left(p\frac{e^{-|k|\sqrt{p^2-1}}}{\sqrt{p^2-1}}-e^{-p|k|}\right),
\end{equation}
the inverse Laplace transform of which is \cite[p. 64, Eq. 15]{Prudnikov}
\begin{equation}
    F(\zeta) =\frac{\zeta}{2\sqrt{\zeta^2-k^2}}I_{1}(\sqrt{\zeta^2-k^2}).
\end{equation}
Using in addition that $\int_{-\infty}^{\infty} \frac{dy}{2\pi}\cos(y)\cos(sy)=\frac{\delta(s-1)+\delta(s+1)}{2}$, we finally obtain Eq.~\eqref{eq:Zconstscal}.

\subsection{Constraint correlation function}

Starting from Eq.~\eqref{eq:GG} and using Eq.~\eqref{eq:phi} with $h=i\theta$, we remark that the integral is again invariant under the change $\theta\to-\theta$ and $S\to-S$. For the same reason as given for the constraint partition function, the integrand is also periodic of period $\pi$, which allows for writing Eq.~\eqref{eq:GGtheta}.

In the scaling limit $N\to\infty$ keeping $Ne^{-2K}=\zeta$, $S/N=s$ and $x=r/N$ constant, $\mathbf{g}_\zeta(x;s)=\lim_{N\to\infty} \frac{N}{2}\mathbf{G}_N(r;S)$ becomes
\begin{equation}
\begin{split}
    \mathbf{g}_\zeta(x;s)&=\lim_{N\to\infty}N\int_{-\pi/2}^{\pi/2} \frac{d\theta}{2\pi} \bigg(\frac{\lambda_+(i\theta)^{N(1-x)}\lambda_-(i\theta)^{Nx}+\lambda_-(i\theta)^{N(1-x)}\lambda_+(i\theta)^{Nx}}{\mathcal Z_N(0)}\sin^2(\phi(i\theta))\\
     &\quad\quad\quad\quad\quad\quad\quad\quad\quad\quad\quad    +\frac{\lambda_+(i\theta)^{N}+\lambda_-(i\theta)^{N}}{\mathcal Z_N(0)}\cos^2(\phi(i\theta))\bigg)\cos(Ns\theta),\\
                    &= \int_{-\infty}^{\infty}\frac{dy}{2\pi}\cos(sy) \left(\frac{\zeta^2}{\zeta^2-y^2}\frac{\cosh\left((1-2x)\sqrt{\zeta^2-y^2}\right)}{\cosh(\zeta)}-\frac{y^2}{\zeta^2-y^2}\frac{\cosh(\sqrt{\zeta^2-y^2})}{\cosh(\zeta)}\right).
\end{split}
\end{equation}

To compute $\int_0^1dx\,\mathbf{g}_\zeta(x;s)$, we use
\begin{equation}
\begin{split}
    \int_0^1dx\,\mathbf{g}_\zeta(x;s)
                    &= \int_{-\infty}^{\infty}\frac{dy}{2\pi}\cos(sy) \left(\frac{\zeta^2}{(\zeta^2-y^2)^{3/2}}\frac{\sinh\left(\sqrt{\zeta^2-y^2}\right)}{\cosh(\zeta)}-\frac{y^2}{\zeta^2-y^2}\frac{\cosh(\sqrt{\zeta^2-y^2})}{\cosh(\zeta)}\right),\\
                     &= -\int_{-\infty}^{\infty}\frac{dy}{2\pi}\cos(sy) \frac{\partial^2}{\partial y^2}\left(\frac{\cosh\left(\sqrt{\zeta^2-y^2}\right)}{\cosh(\zeta)}\right),\\
                     &=s^2\int_{-\infty}^{\infty}\frac{dy}{2\pi}\cos(sy)\frac{\cosh\left(\sqrt{\zeta^2-y^2}\right)}{\cosh(\zeta)},\\
                     &=s^2 z_\zeta(s).
\end{split}
\end{equation}

Finally, the Fourier transform of $\mathbf{g}_\zeta(x;s)$ can be computed ($n\neq0$, $\tilde y=y/\zeta$ and $k=s\zeta$),
\begin{equation}
\begin{split}
    \mathbf{g}_\zeta(n;s) &= \int_0^1 dx \int_{-\infty}^{\infty}\frac{dy}{2\pi} \frac{\zeta^2}{\zeta^2-y^2}\frac{\cosh((1-2x)\sqrt{\zeta^2-y^2})}{\cosh(\zeta)}2\cos(2\pi n x)\cos(sy),\\
                            &=\frac{\zeta}{\cosh(\zeta)}\int_{-\infty}^{\infty}\frac{d\tilde y}{2\pi}\frac{\cos(k\tilde y)}{1-\tilde y^2}\int_0^1 dx2\cos(2\pi n x)\cosh((1-2x)\zeta\sqrt{1-\tilde y^2}),\\
                            &=\frac{2}{\cosh(\zeta)}\int_{-\infty}^{\infty}\frac{d\tilde y}{2\pi}\frac{\cos(k\tilde y)}{\sqrt{1-\tilde y^2}}\frac{\cosh(\zeta\sqrt{1-\tilde y^2})}{1-\tilde y^2+b^2},\\
                            &=\frac{F_1(\zeta)}{\cosh(\zeta)},
\end{split}
\end{equation}
with $b=n\pi/\zeta$ and the last line defines $F_1(\zeta)$.

As above, we compute the Laplace transform of $F_1(\zeta)\to \hat F_1(p)$, and then perform the integral over $\tilde y$, which gives
\begin{equation}
    \hat F_1(p) = \frac{1}{b^2+p^2}\left(\frac{e^{-|k|\sqrt{p^2-1}}}{\sqrt{p^2-1}}-i\frac{e^{i|k|\sqrt{1+b^2}}}{\sqrt{1+b^2}}\right).
\end{equation}
The inverse Laplace transform of the second term in the right-hand-side is proportional to $\sin(\zeta b)=\sin(n\pi)$ and thus vanishes. Knowing that the Laplace transform of $\sin(\zeta b)/b$ is $(p^2+b^2)^{-1}$ and that of 
$\Theta(\zeta-|k|)I_0(\sqrt{\zeta^2-|k|^2})$ is $\frac{e^{-|k|\sqrt{p^2-1}}}{\sqrt{p^2-1}}$, with $\Theta$ the Heaviside function, \cite[p. 62, Eq. 8]{Prudnikov}, and using the convolution properties of Laplace transform, we finally get 
\begin{equation}
\begin{split}
    \mathbf{g}_\zeta(n;s) & = \frac{\zeta^2}{\cosh(\zeta)}\int_{|s|}^1 dt\frac{\sin((1-t)n\pi)}{n\pi} I_0(\zeta\sqrt{t^2-s^2}).
\end{split}
\end{equation}

The real-space correlation function is given by
\begin{equation}
    \mathbf{g}_\zeta(x;s)=s^2 z_\zeta(s)+\sum_{n>0}\mathbf{g}_\zeta(n;s)\cos(2\pi n x),
\end{equation}
where the first term in the right-hand side corresponds to the mode $n=0$. Using $\sum_{n>0}\sin(2\pi n x)/\pi n=f(x)$, with $f(x+1)=f(x)$ and $f(x)=\dfrac{1}{2}-x$ for $x\in ]0,1[$, we find
\begin{equation}
    \sum_{n>0}\cos(2\pi n x)\frac{\sin((1-t)\pi n}{n\pi}=\frac12\left(t-\Theta\left(x-\frac{1-t}{2}\right)\Theta\left(\frac{1+t}{2}-x\right)\right),
\end{equation}
which leads to
\begin{equation}
    \mathbf{g}_\zeta(x;s)=s^2 z_\zeta(s)+\frac{\zeta^2}{2\cosh(\zeta)}\int_{|s|}^1 dt\left(t-\Theta\left(x-\frac{1-t}{2}\right)\Theta\left(\frac{1+t}{2}-x\right)\right) I_0\left(\zeta\sqrt{t^2-s^2}\right).
\end{equation}

This expression can be simplified further using
\begin{equation}
   \int_{|s|}^1 dt\,  t \,I_0\left(\zeta\sqrt{t^2-s^2}\right)=\frac{\sqrt{1-s^2}}{\zeta}I_1\left(\zeta\sqrt{1-s^2}\right),
\end{equation}
which implies
\begin{equation}
    \mathbf{g}_\zeta(x;s)= z_\zeta(s)-\frac{\zeta^2}{2\cosh(\zeta)}\int_{{\rm max}(|s|,|1-2x|)}^1 I_0\left(\zeta\sqrt{t^2-s^2}\right) dt.
\end{equation}

\section{Constraint correlation function in the presence of two domain walls\label{app:domain}}

We compute here the constraint correlation function for $|s|<1$ at very low temperatures corresponding to the limit $N\ll \xi$, i.e. $\zeta\to 0$.
For this, we first remark that the energy cost of a domain wall between domains of up and down spins is $2K$. Furthermore, due to the periodic boundary conditions, there is always an even number of domain walls. By imposing a total spin $S$, we force the system to have $N_\uparrow = \frac{N+S}{2}$ spins up and $N_\downarrow = \frac{N-S}{2}$ spins down. As shown in \cite{Antal2004}, the probability of having $2m$ domain walls goes like $e^{-4Km}$ while the number of possible configurations to place the domains is 
\begin{equation}
    \Omega_m(S)=\begin{pmatrix}
        N_\uparrow-1 \\
        m-1
    \end{pmatrix}\begin{pmatrix}
        N_\downarrow \\
        m
    \end{pmatrix}
    +
    \begin{pmatrix}
        N_\downarrow-1 \\
        m-1
    \end{pmatrix}\begin{pmatrix}
        N_\uparrow \\
        m
    \end{pmatrix}.
\end{equation}
Thus, the total weight for having $2m$ domain walls when the total spin is $S$ reads 
\begin{equation}
    W_m(S) = \frac{e^{-4Km}}{(2\cosh(K))^N} \Omega_m(S),
\end{equation}
such that $Z_N(S)=\sum_{m=0}^\infty W_m(S)$. In the scaling limit, $N\to\infty$ keeping $Ne^{-2K}=\zeta$, $S/N=s$, with $n_\uparrow=\frac{1+s}{2}$ and $n_\downarrow=\frac{1-s}{2}$, $w_m(s)=\frac{N}{2}W_m(S)$ is given by
\begin{equation}
    w_m(s) = \frac{\zeta^{2m}}{4\cosh(\zeta)} \frac{[(1-s^2)/4]^{m-1}}{m!(m-1)!},
\end{equation}
such that $z_\zeta(s) = \sum_{m=0}^\infty w_m(s)$ for $|s|\neq 1$ (the two Dirac deltas need to be treated separately). In the limit $\zeta\to 0$, the dominant term thus comes from the $N$ two-domain-wall-configurations, with $Z_N(S)\simeq N e^{-4K}/(2\cosh(K))^N$ or $z_\zeta(s)\simeq \zeta^2/4\cosh(\zeta)$.

To compute $G_N(r;S)=\langle \sigma_i\sigma_j\delta_{\sum_i\sigma_i,S} \rangle/Z_N(S)$ in the limit of two domain walls, we label the position of the leftmost spin of the up-domain as $l$, while the position of the leftmost spin of the down-domain is $l+N_\uparrow$. Then the value of $\sigma_i$ is $+1$ if $i\in\{l,\ldots,l+N_\uparrow-1\}$ and $-1$ else, and similarly for $\sigma_j$, and we should sum over all possible $l$.

Instead, we prefer to perform a change of frame, and fix $l=1$ and sum over all possible $i$ with $j=i+r$. Then $\sigma_i={\rm sign}(N_\uparrow-i)$ and $\sigma_j={\rm sign}(N_\uparrow-i-r){\rm sign}(N-i-r)$, with the convention that ${\rm sign}(0)=1$. The second ${\rm sign}$ in $\sigma_j$ is needed to implement the periodic boundary conditions when $i+r> N$.

Thus, in this limit, we find that 
\begin{equation}
    G_N(r;S)= \frac{1}{N}\sum_{i=1}^{N}{\rm sign}(N_\uparrow-i){\rm sign}(N_\uparrow-i-r){\rm sign}(N-i-r)
\end{equation}
An explicit calculation gives 
\begin{equation}
\begin{split}
    G_N(r;S)=\begin{cases}
			1-4r/N, & \text{if $0\leq r\leq \frac{N-|S|}{2}$},\\
                2|S|/N-1, & \text{if $\frac{N-|S|}{2}\leq r\leq \frac{N+|S|}{2}$},\\
                4r/N-3, & \text{if $\frac{N+|S|}{2}\leq r\leq N$},
		 \end{cases}
\end{split}
\end{equation}
In the scaling regime, $N\to\infty$ with $x=r/N$ constant, we recover Eq.~\eqref{eq:g2domain}.

\let\oldbibliography\thebibliography
\renewcommand{\thebibliography}[1]{%
  \oldbibliography{#1}%
  \setlength{\baselineskip}{10pt}
  \setlength{\lineskiplimit}{-\maxdimen}
}
\bibliography{biblio}

\end{document}